\documentclass[aps,prd,12pt,notitlepage,showpacs]{revtex4-1}
\usepackage{graphicx}
\usepackage{amsmath}
\usepackage{amssymb}
\usepackage{mathrsfs}
\usepackage{amsfonts}
\usepackage{dsfont}
\usepackage{dcolumn}
\usepackage{bm}
\usepackage{booktabs}
\usepackage{textcomp}
\usepackage{color}
\usepackage{xcolor}
\usepackage[retainorgcmds]{IEEEtrantools}
\usepackage{bigints}

\providecommand{\abs}[1]{\left\lvert#1\right\rvert}
\providecommand{\pd}[2]{\frac{\partial#1}{\partial#2}} 
\providecommand{\pdd}[2]{\frac{\partial^2#1}{\partial#2^2}}
\providecommand{\fracd}[2]{\frac{\displaystyle{#1}}{\displaystyle{#2}}}

\newcommand{\mL}{\mathcal{L}}

\newcommand{\bep}{\bm{\epsilon}}
\newcommand{\br}{\mathbf{r}}
\newcommand{\vx}{\mathbf{x}}

\newcommand{\vp}{\mathbf{p}}

\newcommand{\bpsi}{{\psi^*}}

\newcommand{\bnabla}{\bm{\nabla}}

\usepackage[all]{xy}

\begin{document}

\setlength{\baselineskip}{.85\baselineskip}
%
%

\title{\large Field theory of monochromatic  optical beams. I \\ {\textsl{Classical fields}}}
\author{Andrea Aiello}\email{andrea.aiello@mpl.mpg.de}
\affiliation{Max Planck Institute for the Science of Light,\\
G\"{u}nter-Scharowsky-Stra{\ss}e 1/Bau 24, 91058 Erlangen, Germany}
\affiliation{Institute for Optics, Information and Photonics, \\
 University of Erlangen-Nuernberg, Staudtstrasse 7/B2, 91058 Erlangen, Germany}

\begin{abstract}
We study monochromatic, scalar solutions of the Helmholtz and paraxial wave equations  from a field-theoretic point of view.
We introduce appropriate time-independent Lagrangian densities for which the Euler-Lagrange equations reproduces either Helmholtz and paraxial wave equations with the $z$-coordinate, associated with the main direction of propagation of the fields, playing the same role of time in standard Lagrangian theory. For both Helmholtz and paraxial scalar fields, we calculate the canonical energy-momentum tensor  and determine the continuity equations relating ``energy'' and ``momentum'' of the fields.
Eventually, the reduction of the Helmholtz wave equation to a useful first-order Dirac form, is presented. 
 This work sheds some light on the intriguing and not so acknowledged connections between angular spectrum representation of optical wavefields, cosmological models and physics of black holes. 
\end{abstract}
\pacs{xx.xx.Aa}

\date{\today}
\maketitle

\section{Introduction}

Light is an electromagnetic phenomenon which can be described by a field theory governed by Maxwell equations. These are a set of first-order partial differential equations that relates electric and magnetic \emph{vector} fields each other and  to external sources when present.  However, in many practical instances, a  vector field representation of light appears redundant and a simpler \emph{scalar} field description results appropriate.  In these cases, according to the characteristics of the phenomenon under investigation, \emph{monochromatic light} propagating in free space can be described either by a field $\psi(\vx,z)=\psi(x,y,z)$ obeying the Helmholtz wave equation (HWE)
\begin{align}
\left( \frac{\partial^2}{ \partial x^2} + \frac{\partial^2}{ \partial y^2} + \frac{\partial^2}{ \partial z^2} + k^2_0 \right)\psi(\vx,z) = 0, \qquad k_0>0,\label{f10}
\end{align}
or by  a field $\phi(\vx,z)=\phi(x,y,z)$ satisfying the paraxial wave equation (PWE)
\begin{align}
\left( \frac{\partial^2}{ \partial x^2} + \frac{\partial^2}{ \partial y^2} + 2 i k_0 \frac{\partial}{ \partial z}  \right)\phi(\vx,z) = 0, \qquad k_0>0,\label{f20}
\end{align}
with $\vx = (x,y) \in \mathbb{R}^2$.

In the appendix XI of their book ``Principles of Optics'', Born and Wolf   derive the  energy conservation law for a \emph{real}, time-dependent scalar wavefield $V^{(r)}(\mathbf{r},t)$ in free space, with $\br = (x,y,z) \in \mathbb{R}^3$ \cite{BW}. Because of the explicit time dependence of $V^{(r)}(\mathbf{r},t)$,  a continuity equation expressing the local energy conservation law could be deduced from the Lagrangian form of field equations. For the case of a monochromatic field of frequency $\omega$, Born and Wolf first rewrite the real field as $V^{(r)}(\mathbf{r},t) = \mathcal{R} \{U(\mathbf{r}, \omega)e^{- i \omega t} \}$, where $\mathcal{R}$ denotes the real part. Then, they  take the time averages of the energy density and the energy flux vector to obtain conservation laws involving only the time-independent \emph{complex} field $U(\mathbf{r},\omega)$. 

In this work we pursue the same goal of Born and Wolf, yet following an entirely different and new approach. Instead of considering time-dependent monochromatic fields and erasing such dependence via time averages, we develop an \emph{ab initio} time-independent theory  taking  the monochromatic Helmholtz and paraxial wave equations \eqref{f10} and \eqref{f20} as the central points around which we build a time-independent Lagrangian field theory. The idea is to deal with \emph{action functionals} of the form 
\begin{align}
S =   \int_{z_1}^{z_2} d \mathbf{r} \,\mL,\label{f30}
\end{align}
where  $z_1$ and $z_2$ are the limits of integration for the variable $z$ which is associated with the main propagation direction of the field,  $d \mathbf{r} = dx \, dy \, dz$ is the volume measure and $\mL$ denotes the Lagrange density (Lagrangian, for short). 
We require $S$ to be stationary for arbitrary variations of the field quantities that vanish at the end points, namely
\begin{align}
\delta S =   0,\label{f40}
\end{align}
in order to infer the Euler-Lagrange equations reproducing \eqref{f10} and \eqref{f20}. Thus, in our \emph{nonstandard} approach,  propagation along the $z$-axis of a time-independent field obeying either HWE or PWE, is  formally described in the same manner time evolution of a time-dependent field is depicted in the \emph{standard} Lagrangian formalism.

\section{Nonstandard Lagrangian formalism for Helmholtz fields}


In this section we discuss the classical mechanics of a
 \emph{complex} scalar  field  $\psi=\psi(\vx,z)$, which is a solution of the Helmholtz wave equation \eqref{f10}
\begin{align}
\left( \partial^2 + k_0^2 \right)\psi=0,\label{f50}
\end{align}
where the Laplacian $\partial^2$ in the $3$-dimensional Euclidean space $\mathbb{R}^3$ is written as  (we always use the summation convention)
\begin{align}
\partial^2 = \partial_\mu \partial^\mu = g^{\mu \nu} \partial_\mu \partial_\nu, \qquad \left(\mu, \nu = 1,2,3 \right),\label{f60}
\end{align}
with $ g^{\mu \nu} =  \delta^{\mu \nu}$. The three-dimensional gradient is expressed as $\partial \psi/ \partial x^\mu \equiv \partial_\mu \psi = \left(\bm{\nabla},\partial_z \right) \psi$, where a point in $\mathbb{R}^3$ is labeled by the three coordinates $x^\mu$, with $x^3 = z$ the \emph{longitudinal} coordinate and $x^k, \; k=1,2$ the \emph{transverse} coordinates.
The \emph{two-dimensional} gradient of a scalar function $f(x,y,z)$ is denoted $\bm{\nabla} f$ and is defined as
\begin{align}
\bm{\nabla} f = \frac{\partial f}{\partial x} \bep_1 + \frac{\partial f}{\partial y} \bep_2,\label{f65}
\end{align}
where   $\bep_1$ and $\bep_2$ are the orthogonal unit vectors pointing in the $x$ and $y$ Cartesian coordinate directions, respectively.
From now on, Greek indexes $\mu, \nu,\alpha,\beta, \ldots$ , run from $1$ to $3$, while Latin indexes $i,j,k,l,m,n, \ldots$ , take the values $1$ and $2$. 

When the complex scalar field $\psi$ has two \emph{independent} real components $\psi_1$ and $\psi_2$, we may put
\begin{align}
\psi = & \; \left(\psi_1 + i \psi_2 \right)/\sqrt{2}, \label{f70} \\
\psi^* = & \; \left(\psi_1 - i \psi_2 \right)/\sqrt{2}, \label{f80} 
\end{align}
and regard to $\psi$ and $\psi^*$ (instead of $\psi_1$ and $\psi_2$) as independent fields.
In this case  we expect that the two Euler-Lagrange equations 
\begin{align}
\frac{\partial \mL}{\partial \bpsi} - \frac{\partial}{\partial x^\mu} 
\frac{\partial \mL}{\partial (\partial_\mu \bpsi)} = & \; 0,   \label{f90} \\ 
\frac{\partial \mL}{\partial \psi} - \frac{\partial}{\partial x^\mu} 
\frac{\partial \mL}{\partial (\partial_\mu \psi)} = & \; 0,  \label{f100} 
\end{align}
 give the two Helmholtz equations
\begin{align}
\left( \partial^2 + k_0^2 \right)\psi=0, \label{f110}  \\ 
\left( \partial^2 + k_0^2 \right)\bpsi=0,  \label{f120} 
\end{align}
respectively, when an appropriate Lagrangian $\mL = \mL_\text{HWE}$ is chosen.
Thus, our goal now is to find a proper Lagrangian $\mL_\text{HWE}$. To this end, suppose that 
\begin{align}
\mL_\text{HWE} = A \, \delta^{\mu \nu}\partial_\mu \bpsi  \partial_\nu \psi + B \, \bpsi \psi,\label{f130}
\end{align}
where $A,B$ are real constants to be determined. Substitution of Eq. \eqref{f130} into Eq. \eqref{f90} gives
\begin{align}
B \, \psi - A \, \partial^2 \psi = 0,\label{f140}
\end{align}
which coincides with \eqref{f110} if one chooses  $B = - A \, k_0^2$ ($A $ remains undetermined, therefore we are free to choose $A=1$). Thus, the sought Lagrangian is 
\begin{align}
\mL_\text{HWE} =  \delta^{\mu \nu}\partial_\mu \bpsi  \partial_\nu \psi -  k_0^2 \, \bpsi \psi.\label{f150}
\end{align}
From the action principle, it follows that is always possible to add a three-divergence to the Lagrangian \eqref{f150} without altering the equation of motion \eqref{f110}. Therefore, since
\begin{align}
\partial_\mu \left(\delta^{\mu \nu}\bpsi  \partial_\nu \psi \right) = \delta^{\mu \nu}\partial_\mu \bpsi  \partial_\nu \psi + \bpsi \partial^2 \psi,\label{f160}
\end{align}
then we can rewrite Eq. \eqref{f150} in the following \emph{equivalent} form:
\begin{align}
\mL_\text{HWE} = \bpsi \left( \partial^2 + k_0^2 \right) \psi.\label{f170}
\end{align}
The Euler-Lagrange equation \eqref{f90} now simply becomes
\begin{align}
\frac{\partial \mL}{\partial \bpsi} = \left( \partial^2 + k_0^2 \right) \psi = 0.\label{f180}
\end{align}

\subsection{Structural aspects}

Equation \eqref{f110} admits \emph{separable} solutions of the form
\begin{align}
\psi(\vx,z) = \varphi(\vx) e^{i \zeta z},\label{f190}
\end{align}
where by definition $\varphi(\vx)$ satisfies 
\begin{align}
\zeta^2 \varphi(\vx) = \left( \nabla^2 + k_0^2 \right)\varphi(\vx) .\label{f200}
\end{align}
Elementary examples thereof are given by plane wave fields $\psi(\vx,z) = \exp\left(i \vp \cdot \vx \right)\exp[i z (k_0^2 - \vp \cdot \vp)^{1/2} ]$, with  $\vp = (p_1,p_2)\in \mathbb{R}^2$, and by Bessel fields $J_0(k_0 \sin \vartheta_0 (x^2 + y^2)^{1/2}) \exp(i z k_0 \cos \vartheta_0 )$, where $J_0(z)$ denotes the zeroth-order Bessel function of first kind.
The ``frequency'' $\zeta$ can be either real and positive, or \emph{purely imaginary}, namely $\zeta^* = - \zeta$ and $\zeta^2 <0$. This can be seen multiplying Eq. \eqref{f200} by $ \varphi^*(\vx)$ and integrating over the $xy$-plane, thus obtaining \cite{Kang96}
\begin{align}
\zeta^2  = \fracd{k_0^2 \int d \vx \,\abs{\varphi(\vx)}^2 - \int d \vx \,\abs{\bm{\nabla}\varphi(\vx)}^2}{\int d \vx \, \abs{\varphi(\vx)}^2} ,\label{f210}
\end{align}
where $d \vx = d x \, dy$ is the surface element and we assumed that the field $ \varphi(\vx)$ vanishes for $x,y \to \infty$ in order to neglect surface terms.  From Eq. \eqref{f210} it follows that either $\zeta^2 \geq 0$, or  $\zeta^2 <0$ whenever 
\begin{align}
\int d \vx \,\abs{\bm{\nabla}\varphi(\vx)}^2 > k_0^2 \int d \vx \,\abs{\varphi(\vx)}^2.\label{f220}
\end{align}
This relation imposes a constraint upon the Fourier spectrum $\widetilde{\varphi}(\vp)$ of the field $\varphi(\vx)$, where  $\vp = (p_1,p_2)\in \mathbb{R}^2$. Substituting the Fourier representation 
\begin{align}
\varphi(\vx) = \frac{1}{2 \pi} \int d \vp \,\widetilde{\varphi}(\vp) e^{- i \vp \cdot \vx}\label{f230}
\end{align}
into Eq. \eqref{f220} yields to
\begin{align}
\int d \vp \left( p^2 - k_0^2 \right) \abs{\widetilde{\varphi}(\vp)}^2 > 0,\label{f240}
\end{align}
where $p^2 = p_i p_i = \vp \cdot \vp$. Therefore, whenever the support of $\widetilde{\varphi}(\vp)$ is not entirely contained within the circle of equation $p^2 = p_1^2 + p_2^2 = k_0^2$, the field develops purely imaginary frequencies. This fact will have profound consequences upon the quantization of $\psi(\vx,z)$. Since the right side of Eq. \eqref{f210} is always 
\emph{real}, it follows that
\begin{align}
\zeta^2  = \left( \zeta^2 \right)^* = \left( \zeta^* \right)^2 .\label{f215}
\end{align}
Using this result in Eq. \eqref{f200} shows that $\varphi(\vx)$ and $\varphi^*(\vx)$ satisfy the \emph{same} equation. Therefore, if $\varphi(\vx)$ is a given solution of Eq. \eqref{f200}, then  $\varphi^*(\vx)$ is also a solution.

We conclude this part by noticing that, irrespective of the either positive or purely imaginary value taken by $\zeta$, there are four linearly independent separable solutions of Eq. \eqref{f110}, namely
\begin{align}
\psi_+ = \varphi(\vx) e^{ i \zeta z}, \qquad  \psi_+^* = \varphi^*(\vx) e^{ -i \zeta^* z},  \qquad \psi_-^* =  \varphi^*(\vx) e^{ i \zeta^* z},  \qquad \psi_- = \varphi(\vx) e^{- i \zeta z}.\label{f250}
\end{align}
When $\zeta = \zeta^*$  all solutions are oscillatory and therefore physically acceptable. However, when $\zeta  = i \abs{\zeta}$ it has
\begin{align}
\psi_+ = \varphi(\vx) e^{ - \abs{\zeta} z}, \qquad  \psi_+^* = \varphi^*(\vx) e^{ - \abs{\zeta} z},  \qquad \psi_-^* =  \varphi^*(\vx) e^{   \abs{\zeta} z},  \qquad \psi_- = \varphi(\vx) e^{ \abs{\zeta} z}\label{f260}
\end{align}
and the $\psi_-$'s solutions are \emph{exponentially} growing as $z$ increases. Therefore they represent physically acceptable solutions only for $z<0$. Vice versa, the $\psi_+$'s solutions are  exponentially decaying and physically acceptable only for $z>0$. For fields associated to optical beams, $\psi_+$ ($\psi_-$) and $\psi_+^*$  ($\psi_-^*$) are called \emph{evanescent waves} when $\zeta  = i \abs{\zeta}$ and $z>0$ ($z<0$).

\subsection{Symmetries and conservation laws}

In this part we discuss the symmetries of the Lagrangian \eqref{f150}. To begin with, let us note that
such Lagrangian is manifestly invariant under the transformation
\begin{align}
\psi \to e^{- i \Lambda} \psi, \qquad \bpsi \to e^{i \Lambda} \bpsi,\label{f270}
\end{align}
where $\Lambda$ is a real constant. From the Noether's theorem follows that there exist a current 
\begin{align}
\mathscr{J}_\mu = & \; \frac{1}{i} \left[ \psi \pd{\mL}{(\partial^\mu \psi)} - \bpsi \pd{\mL}{(\partial^\mu \bpsi)} \right] \nonumber \\
= & \; \frac{1}{i} \bigl( \psi \partial_\mu \bpsi - \bpsi \partial_\mu \psi \bigr),\label{f280}
\end{align}
which has a vanishing three-divergence \cite{Ryder,LowellBrown,Weinberg}
\begin{align}
\partial^\mu \mathscr{J}_\mu = \partial_z \mathscr{J}_z + \bnabla \cdot \pmb{\mathscr{J}}=0,\label{f285}
\end{align}
namely
\begin{align}
\pd{}{z} \left( \psi \pd{ \bpsi}{z} - \bpsi \pd{ \psi}{z} \right) = - \bnabla \cdot \bigl( \psi \bnabla \bpsi - \bpsi \bnabla \psi \bigr)\label{f288}
\end{align}
%
%
%
%
%
%
Integrating both sides of this equation over all the $xy$-plane we obtain 
\begin{align}
\partial_z \int d \vx \mathscr{J}_z = & \;  - \int d \vx \, \bnabla \cdot \pmb{\mathscr{J}} \nonumber \\
= & \; 0 ,\label{f320}
\end{align}
where the right side amounts to the two-dimensional integral of a two-divergence
 and then vanishes for fields localized within a \emph{finite} region of the $xy$-plane.
This equation states that during propagation of a monochromatic  optical field along the $z$-axis,  the ``charge'' $Q$ defined as
\begin{align}
Q = \frac{1}{i} \int d \vx \, \left( \psi \pd{ \bpsi}{z} - \bpsi \pd{ \psi}{z} \right), \qquad \text{is conserved:} \qquad \pd{Q}{z}=0.\label{f330}
\end{align}
It is instructive to evaluate $Q$ for the four fundamental solutions \eqref{f250}. A straightforward calculation gives
\begin{align}
\psi_+ \to  & \; Q_+ = - 2  \, C \, e^{- 2 z \operatorname{Im} \zeta} \operatorname{Re} \zeta,   \label{f335a} \\ 
\psi_+^* \to  & \; -Q_+ ,   \label{f335b} \\ 
\psi_-^* \to  & \; -Q_- ,   \label{f335c} \\ 
\psi_- \to  & \; Q_- =  2 \, C \, e^{2 z \operatorname{Im} \zeta} \operatorname{Re} \zeta,   \label{f335d} 
\end{align}
where 
\begin{align}
C = \int d \vx \, \abs{\varphi(\vx)}^2.\label{f337}
\end{align}
At first sight, the charges $Q_\pm$ seems to depend on $z$, thus contradicting the conservation law \eqref{f330}. However, one should remember that there are only two possibilities for $\zeta$: either $\zeta=  \zeta^* \Rightarrow \operatorname{Im} \zeta =0$, or $\zeta= - \zeta^* \Rightarrow \operatorname{Re} \zeta =0$. In the first case we have $Q_\pm = \mp 2 C \zeta$ and the conservation law \eqref{f330} is satisfied. In the latter case $Q_+ = 0 = Q_-$ and there is no conserved charge. Thus, there is no charge associated to the evanescent waves.

Additional conserved quantities can be calculated in a straightforward manner from the \emph{canonical energy-momentum tensor}, which is calculated from the Lagrangian as
\begin{align}
\mathscr{T}_{\mu \nu} = & \; \pd{\mL}{(\partial^\mu \psi)} \partial_\nu \psi + \pd{\mL}{(\partial^\mu \bpsi)} \partial_\nu \bpsi - \delta_{\mu \nu} \mL \nonumber \\
= & \; \partial_\mu \bpsi \partial_\nu \psi + \partial_\mu \psi \partial_\nu \bpsi - \delta_{\mu \nu} \bigl( \delta^{\alpha \beta} \partial_\alpha \bpsi \partial_\beta \psi - k_0^2 \bpsi \psi \bigr).\label{f340}
\end{align}
Note that this tensor is automatically symmetric, namely $\mathscr{T}_{\mu \nu} = \mathscr{T}_{ \nu \mu}$ and, by definition,
\begin{align}
\pd{}{x_\mu}\mathscr{T}_{\mu \nu} = 0.\label{f350}
\end{align}
Since $\nu =1,2,3$ this means that there are three conserved quantities,  the ``energy'' $H$ and the linear momentum vector  $\mathbf{P}$ of the field, defined respectively as
\begin{align}
H =  \int d \vx \, \mathscr{T}^{33} = \int d \vx \, \left(\abs{\partial_z \psi}^2 - \abs{\bnabla \psi}^2 + k_0^2 \abs{\psi}^2 \right), \label{f360}
\end{align}
and 
\begin{align}
\mathbf{P} =  \int d \vx \, \mathscr{T}^{3 l} \bep_l  =  \int d \vx \, \left( \pd{\, \bpsi}{z} \bnabla \psi + \pd{\, \psi}{z} \bnabla \bpsi \right),\label{f370}
\end{align}
where $P^\mu = \left(H, \mathbf{P} \right)$ denotes the full three-momentum of the beam.
Therefore, as the beam propagates along the $z$-axis, the ``energy'' $H$ and the linear momentum $ \mathbf{P}$ remain constant as a consequence of Eq. \eqref{f350}, namely 
\begin{align}
\pd{H}{z} = 0 = \pd{\mathbf{P}}{z}.\label{f380}
\end{align}

As it will be shown later, the ``energy'' $H$ coincides with the Hamiltonian of the system. However, the expression in Eq. \eqref{f360} is not manifestly \emph{positive semidefinite}, as a physically realizable Hamiltonian should be, because of the negative ``kinetic energy'' term $- \abs{\bnabla \psi}^2$. We will discuss this point at length  later, when proceeding with the quantization of the field. For the moment, we verify that $H$ is actually positive semidefinite 
 for the four fundamental solutions \eqref{f250}. After a straightforward calculation one finds 
\begin{align}
\psi_\pm  \to  H = & \; e^{\mp 2 z \operatorname{Im} \zeta} \int d \vx \, 
\left( \abs{\zeta}^2 \abs{\varphi(\vx)}^2 - \abs{\bnabla \varphi(\vx)}^2 + k_0^2 \abs{\varphi(\vx)}^2\right) \nonumber \\
= & \; \left(\zeta^2 + \abs{\zeta}^2 \right) e^{\mp 2 z \operatorname{Im} \zeta} \int d \vx \, \abs{\varphi(\vx)}^2,\label{f390} 
\end{align}
where we used the equation of motion Eq. \eqref{f200} and integration by part (discarding surface terms) to pass from the first to the second expression. The conjugate fields $\psi_\pm^*$ yield the same  $H$. Again, Eq. \eqref{f390} seems to depend upon the propagation distance $z$, but this is not the case because $\zeta^2 + \abs{\zeta}^2 = \zeta(\zeta + \zeta^*) = 2 \, \zeta \operatorname{Re} \zeta$ which implies that
\begin{align}
\begin{cases}
\zeta = \zeta^* & \Rightarrow \quad \zeta^2 = \abs{\zeta}^2 \quad \phantom{xi}\text{when} \quad  \operatorname{Im} \zeta =0, \\
\zeta = -\zeta^* & \Rightarrow \quad \zeta^2 = -\abs{\zeta}^2 \quad \text{when} \quad \operatorname{Re} \zeta =0.\label{f400}
\end{cases}
\end{align}
Therefore, using Eq. \eqref{f400} into Eq. \eqref{f390} we obtain
\begin{align}
\label{f410}
H = \begin{cases}
 2  \, C \, \zeta^2 \geq 0, & \zeta = \zeta^*, \\
0, & \zeta = -\zeta^*,
\end{cases}
\end{align}
where $C$ is given again by Eq. \eqref{f337}. 
This nice result shows that the evanescent waves \emph{do not} contribute to the total energy of the field. 

In a similar manner we can now calculate $\mathbf{P}$ and the outcome is
\begin{align}
\psi_\pm  \to  \mathbf{P} = & \; \pm 2 \operatorname{Re} \zeta \, e^{\mp 2 z \operatorname{Im} \zeta} \, 
\int d \vx \, \varphi^*(\vx) \left( \frac{1}{i} \bnabla \right) \varphi(\vx)   \nonumber \\
= & \; \pm 2 \operatorname{Re} \zeta \, e^{\mp 2 z \operatorname{Im} \zeta} \, 
\int d \vp \,  \vp \abs{\widetilde{\varphi}(\vp)}^2 , \label{f420} 
\end{align}
where the Fourier representation  Eq. \eqref{f230} has been used. The conjugate fields $\psi_\pm^*$ yield the same  $\mathbf{P}$. Also in this case the $z$-dependence of $\mathbf{P}$ is only deceptive because we can always  rewrite Eq. \eqref{f420} as
\begin{align}
\mathbf{P} = \begin{cases}
\displaystyle{\pm 2  \, \zeta  \int d \vp \,  \vp \abs{\widetilde{\varphi}(\vp)}^2}, & \zeta = \zeta^*, \\
0, & \zeta = -\zeta^*.
\end{cases}\label{f415}
\end{align}
Once again, the \emph{physical} transverse linear momentum $\mathbf{P}$ does not receive contributions from the evanescent waves.

Finally, it is of some interest to note that from Eq. \eqref{f350} also follows an interesting \emph{continuity equation} connecting the linear momentum density $\mathscr{P}^i = \mathscr{T}^{3i}$ with the linear momentum \emph{transverse flux density} $\mathscr{T}^{ij}$ \cite{Barnett}:
\begin{align}
\pd{}{z} \mathscr{P}^i + \pd{}{x^j}\mathscr{T}^{ij} = 0, \label{f417}
\end{align}
where 
\begin{align}
\mathscr{T}^{ij} = \partial^i \bpsi \, \partial^j \psi + \partial^j \psi \, \partial^i \bpsi - \delta^{ij}\left( \abs{\partial_\alpha \psi}^2 - k_0^2 \abs{\psi}^2\right). \label{f430}
\end{align}
This equation \eqref{f417} simply tells us that variations of the transverse linear momentum density during possibly occurring during propagation, are compensated by the transverse variations of  $\mathscr{T}^{ml}$. 

Since our canonical energy-momentum tensor is symmetric, we can construct an additional conserved tensor density:
\begin{align}
\mathscr{M}^{\lambda \mu \nu} \equiv  x^\mu \mathscr{T}^{\lambda \nu} - x^\nu \mathscr{T}^{\lambda \mu}, \label{f440}
\end{align}
such that 
\begin{align}
\partial_\lambda \mathscr{M}^{\lambda \mu \nu}  = \mathscr{T}^{\mu \nu} -\mathscr{T}^{\nu \mu} = 0. \label{f450}
\end{align}
To proceed further, we rewrite explicitly Eq. \eqref{f450} as
\begin{align}
\partial_\lambda \mathscr{M}^{\lambda \mu \nu}  = \pd{}{x}\mathscr{M}^{1 \mu \nu} +\pd{}{y}\mathscr{M}^{2 \mu \nu}+\pd{}{z}\mathscr{M}^{3 \mu \nu}= 0. \label{f470}
\end{align}
Then, we integrate term by term over the $xy$-plane obtaining
\begin{align}
\pd{}{z}\int d \vx \, \mathscr{M}^{3 \mu \nu} = - \int d \vx \left(\pd{}{x}\mathscr{M}^{1 \mu \nu}\right) -\int d \vx \left(\pd{}{y}\mathscr{M}^{2 \mu \nu}\right). \label{f475}
\end{align}
%
The two terms at the right side of this equation can be discarded under the assumption that the fields and their derivative fall off sufficiently fast at infinity. Thus, we recover the well-known conservation law \cite{Ryder}:
\begin{align}
\pd{}{z} \int d \vx \,  \mathscr{M}^{3 \mu \nu}  = 0. \label{f460}
\end{align}
However, it should be reminded  that particular care must be taken when handling Eq. \eqref{f475} because of the risk of improper manipulation of the surface terms \cite{Ornigotti14}. 
To proceed further, it  is useful to define  the conserved \emph{angular momentum tensor} as 
\begin{align}
J^{\mu \nu} \equiv \int d \vx \,  \mathscr{M}^{3 \mu \nu}  \equiv  \int d \vx \,  \mathscr{J}^{\mu \nu} , \label{f465}
\end{align}
where 
\begin{align}
\mathscr{J}^{\mu \nu}  = x^\mu \mathscr{P}^{\nu} - x^\nu \mathscr{P}^{\mu} , \label{f480}
\end{align}
with $\mathscr{P}^{\mu} \equiv \mathscr{T}^{3\mu}$. From the definitions above, it follows that $\mathscr{J}^{\mu \nu}$ is antisymmetric, namely  $\mathscr{J}^{\mu \mu} = 0$ (no sum over repeated indices) and $\mathscr{J}^{\mu \nu} = -\mathscr{J}^{\nu \mu}$. Moreover, using the only three independent Cartesian components of $\mathscr{J}^{\mu \nu}$, we define the ``\emph{angular momentum density vector}''  as
\begin{align}
\mathscr{J}_\lambda  = \frac{1}{2} \epsilon_{\lambda \mu \nu} \mathscr{J}^{\mu \nu} = \left( \mathscr{J}^{23}, \mathscr{J}^{31},\mathscr{J}^{12} \right),  \label{f490}
\end{align}
where $ \epsilon_{\lambda \mu \nu}$ denotes the totally antisymmetric three-dimensional Levi-Civita symbol. 
Using Eqs. (\ref{f440},\ref{f480}) into Eq. \eqref{f490} we obtain 
\begin{align}
\mathscr{J}_x = & \; y \mathscr{P}_z - z \mathscr{P}_y,   \label{f500} \\
\mathscr{J}_y = & \; z \mathscr{P}_x - x \mathscr{P}_z,   \label{f510} \\
\mathscr{J}_z = & \; x \mathscr{P}_y - y \mathscr{P}_x.   \label{f520} 
\end{align}
%
%
The corresponding \emph{total} angular momentum components are straightforwardly calculated by integrating the relations above to obtain $J_\lambda  =  \epsilon_{\lambda \mu \nu} J^{\mu \nu}/2$, where Eq. \eqref{f465} has been used. Explicitly, integration both sides of Eqs. (\ref{f500}-\ref{f510}) over the $xy$-plane, gives
\begin{align}
J_x = & \; \overline{y}(z) - z P_y,   \label{fb500} \\
J_y = & \; z P_x - \overline{x}(z),   \label{fb510} \
\end{align}
where we have defined
\begin{align}
\overline{y}(z) \equiv \int d \vx \, \Bigl(y \mathscr{P}_z \Bigr), \qquad \text{and} \qquad \overline{x}(z) \equiv \int d \vx \, \Bigl(x \mathscr{P}_z \Bigr),  \label{ave} 
\end{align}
and Eq. \eqref{f370} has been used. For fields such that $\mathscr{P}_z \geq 0$, the vector $\overline{\vx}(z) = \bep_1 \overline{x}(z) + \bep_2 \overline{y}(z)$ can be interpreted as the \emph{centroid} of the energy distribution on the $xy$-plane. Then, deriving both sides of Eqs. (\ref{fb500}-\ref{fb510}) with respect to $z$ and using the conservation laws  
(\ref{f380},\ref{f460}) we obtain the equations of motions of the centroid of the field
\begin{align}
\frac{d \, \overline{y}(z)}{d  z} = P_y, \qquad \text{and} \qquad \frac{d \, \overline{x}(z)}{d  z} = P_x,  \label{gSHEL} 
\end{align}
which reproduce the laws of \emph{rays} propagation in geometrical optics \cite{gSHEL}.

It is enlightening to calculate explicitly $J_z$ for the fields \eqref{f250}. After a lengthy but straightforward calculation one finds
\begin{align}
\psi_{\pm} \to J_z = \pm 2 \operatorname{Re} \zeta \, e^{\mp 2 z \operatorname{Im} \zeta} \int d \vx \, \varphi^*(\vx) \left( x \frac{1}{i} \pd{}{y} - y \frac{1}{i} \pd{}{x}\right)\varphi(\vx), \label{f522}
\end{align}
where $2 \operatorname{Re} \zeta  \exp (- 2 z \operatorname{Im} \zeta ) = 2 \, \zeta$ for $\zeta = \zeta^*$ and it is equal to zero for $\zeta = -\zeta^*$. Once again, the ``unphysical'' evanescent waves generated by the angular spectrum representation do not carry angular momentum. From Eq. \eqref{f522} it follows  that $J_z$ is conserved along with propagation because $\partial J_z / \partial z \propto \operatorname{Re} \zeta \operatorname{Im} \zeta =0$, the latter equality being a consequence of the fact that $\zeta$ is either real or purely imaginary. The conjugate fields $\psi_\pm^*$ produce the same  $J_z$.

 One more continuity equation may be derived by rewriting Eq. \eqref{f450} with the help of Eq. \eqref{f465}, as 
\begin{align}
\pd{}{z}\mathscr{J}^{\mu \nu} + \pd{}{x^i} \mathscr{M}^{i \mu \nu}=0. \label{f530}
\end{align}
Multiplying both sides of this equation by $\epsilon_{\lambda \mu \nu}/2$ and then summing over repeated indices, we obtain 
\begin{align}
\pd{}{z}\mathscr{J}_{\lambda} + \pd{}{x^i} \mathscr{L}^{i}_{\phantom{s}\lambda} = 0, \label{f540}
\end{align}
where we have defined the \emph{angular momentum flux density} as
\begin{align}
 \mathscr{L}^{\,i}_{\phantom{x}\lambda} \equiv \frac{1}{2} \epsilon_{\lambda \mu \nu} \mathscr{M}^{i \mu \nu}. \label{f550}
\end{align}
This quantity differs from the homonym one introduced by Barnett \cite{Barnett} in that $ \mathscr{L}^{i}_{\phantom{x}\lambda}$ is time-independent.

In conclusion, in this section we have shown that all the relevant physical quantities as energy, linear and angular momenta vanishes for evanescent fields. Thus, the latter appears more as \emph{virtual} fields that do not correspond to any real physical field. However, as we shall see later, they are necessary  to preserve unitarity \cite{Ali72}.

\section{Hamiltonian formalism}

Propaedeutical to the quantization procedure, is the introduction of the Hamiltonian formalism for the Eq. \eqref{f10}. The procedure for passing from the Lagrangian to the Hamiltonian representation of the field is standard \cite{Ryder}. First, we write down our Lagrangian \eqref{f150} as
\begin{align}
\mL = \partial_\mu \bpsi  \partial_\mu \psi -  k_0^2 \, \bpsi \psi,\label{h10}
\end{align}
where, for the sake of clarity we omitted the subscript ``$\text{HWE}$''. Then, we determine the fields $\Pi(\vx,z)$ and $\Pi^*(\vx,z)$ canonically conjugate to $\psi(\vx,z)$ and $\psi^*(\vx,z)$, respectively,
\begin{align}
\Pi = \pd{\mL}{(\partial_3 \psi)} = \partial_3 \psi^*, \qquad \Pi^* = \pd{\mL}{(\partial_3 \psi^*)} = \partial_3 \psi.\label{h20}
\end{align}
The \emph{Hamiltonian density} $\mathscr{H}$ is defined in terms of the four fields $\psi, \Pi, \psi^*,\Pi^*$ as usual:
\begin{align}
\mathscr{H} = & \; \Pi \, \partial_3 \psi + \Pi^* \partial_3 \psi^* - \mL \nonumber \\
= & \; \Pi^* \Pi - \bnabla \psi^* \cdot \bnabla \psi + k_0^2 \psi^* \psi.
\label{h30}
\end{align}
Inverting the relations \eqref{h20} to express the field derivatives in terms of the conjugate momenta, and using this result in Eq. \eqref{h30}, we obtain
\begin{align}
\mathscr{H} = \partial_3 \psi^* \partial_3 \psi - \delta^{ij} \partial_i \psi^* \partial_j \psi + k_0^2 \psi^* \psi,
\label{h40}
\end{align}
which shows that actually $\mathscr{H} = \mathscr{T}^{33}$, as previously stated. The reader familiar with quantum field theory of tachyons, will appreciate the similarity between Eq. \eqref{h40} and the Hamiltonian of a Klein-Gordon field with purely imaginary mass \cite{Schroer,Kamoi}.

Now we are going to show that the canonical Hamiltonian 
\begin{align}
H = \int d \vx \, \mathscr{H} 
\label{h50}
\end{align}
is naturally partitioned in a ``propagating'' and an ``evanescent'' part. Interestingly, the same phenomenon manifests in the quantization of scalar fields  near rapidly rotating stars \cite{Kang96} and in cosmological models of universes with unstable modes \cite{Broad05}. In the latter case the propagating and  evanescent parts are quite suggestively dubbed ``light'' and ``dark'' components of the Hamiltonian, respectively \cite{Broad05}.

At any position $z$ the fields $\Pi(\vx,z)$ and $\psi(\vx,z)$ can be expanded in terms of the Fourier transform representations: 
\begin{align}
\psi(\vx,z) = \frac{1}{2 \pi} \int d \vp \, Q(\vp,z) e^{i \vp \cdot \vx} 
\label{h60}
\end{align}
and 
\begin{align}
\Pi(\vx,z) = \frac{1}{2 \pi} \int d \vp \, P(\vp,z) e^{-i \vp \cdot \vx} ,
\label{h70}
\end{align}
where the complex amplitudes $Q(\vp,z)$ and $P(\vp,z)$ are $z$-dependent. The minus sign in the exponential in Eq. \eqref{h70} is \emph{not} a typo. Substituting Eqs. (\ref{h60}-\ref{h70}) into Eq. \eqref{h50} we obtain, after some manipulation,
\begin{align}
H = \int d \vp \,\Bigl[ P^*(\vp,z)P(\vp,z) + \left(k_0^2 - p^2 \right)Q^*(\vp,z)Q(\vp,z) \Bigr]  ,
\label{h80}
\end{align}
where $p^2 = \vp \cdot \vp$. It is clear that either $\zeta_p^2 \equiv k_0^2 - p^2 \geq 0$ for $  p^2 \leq k_0^2$, or $\zeta_p^2 < 0$ for $  p^2 > k_0^2$. In any case, the $z$-derivatives of $P$ and $Q$ are given by the Hamilton equations
\begin{align}
\pd{}{z} P(\vp,z) = - \frac{ \delta H}{\delta Q(\vp,z )} = - \zeta_p^2 \, Q^*(\vp,z)
\label{h90}
\end{align}
and 
\begin{align}
\pd{}{z} Q^*(\vp,z) =  \frac{ \delta H}{\delta P^*(\vp,z )}  =  P(\vp,z),
\label{h100}
\end{align}
and their conjugates. Here the symbol $\delta F[f]/\delta f[t]$ denotes the functional derivative of the functional $F[f]$ \cite{Ofey}. Deriving Eq. \eqref{h90} with respect to $z$ and using Eq. \eqref{h100} yields the equation of motion of  $Q(\vp,z)$ and $Q^*(\vp,z)$:
\begin{align}
\left( \pdd{}{z} + \zeta_p^2 \right) Q(\vp,z) = 0.
\label{h110}
\end{align}

Before solving this equation, we turn back to Eq. \eqref{h80} to rewrite it as $H = H_L + H_D$, where the subscripts ``$L$'' and ``$D$'' stand for L\emph{ight} and D\emph{ark}, respectively, with
\begin{align}
H_L = \int d \vp \,\Bigl[ P^*(\vp,z)P(\vp,z) + \zeta_p^2 \, Q^*(\vp,z)Q(\vp,z) \Bigr] \Theta(k_0^2 - p^2) ,
\label{h120}
\end{align}
and
\begin{align}
H_D = \int d \vp \,\Bigl[ P^*(\vp,z)P(\vp,z) - \abs{\zeta_p}^2 Q^*(\vp,z)Q(\vp,z) \Bigr] \Theta(p^2 - k_0^2) ,
\label{h130}
\end{align}
where $\Theta(z)$ denotes the Heaviside step function. Equation \eqref{h120} clearly represents the Hamiltonian of a continuum set of harmonic oscillators, because $\Theta(k_0^2 - p^2)\, \zeta_p^2 \geq 0$. However, Eq. \eqref{h130} expresses the Hamiltonian of a continuum set of \emph{repulsive} (or, inverted) harmonic oscillators which are known, in quantum mechanics,  to do not possess neither square-integrable eigenstates, nor  a lower energy vacuum state ($H_D$ is not bounded from below, as we shall see soon) \cite{Barton86}. Therefore, Eq. \eqref{h110} naturally splits in two independent equations of the form
\begin{align}
\left( \pdd{}{z} \pm \abs{\zeta_p}^2 \right) Q_\pm(\vp,z) = 0,
\label{h140}
\end{align}
whose solutions are
\begin{align}
Q_+(\vp,z) = c_{+,1}(\vp)\, e^{i\abs{\zeta_p} z} +  c_{+,2}(\vp)\, e^{-i\abs{\zeta_p} z},
\label{h150}
\end{align}
and 
\begin{align}
Q_-(\vp,z) = c_{-,1}(\vp)\,e^{-\abs{\zeta_p} z} +  c_{-,2}(\vp)\, e^{\abs{\zeta_p} z},
\label{h160}
\end{align}
where $c_{\pm,1}(\vp)$ and $c_{\pm,2}(\vp)$ are arbitrary functions of $\vp$ solely. The corresponding $P_\pm(\vp,z)$ are straightforwardly calculated from Eq. \eqref{h100}: 
\begin{align}
P_+(\vp,z) = & \; \pd{}{z} Q^*_+ (\vp,z) = - i \zeta_p \left( c_{+,1}^*(\vp) \,e^{-i \zeta_p z} -  c_{+,2}^*(\vp)\, e^{i \zeta_p z} \right), \label{h170} \\ 
P_-(\vp,z)= & \; \pd{}{z} Q^*_- (\vp,z) = -\abs{\zeta_p} \left( c_{-,1}^*(\vp) \,e^{-\abs{ \zeta_p} z} -  c_{-,2}^*(\vp)\, e^{\abs{ \zeta_p} z} \right),
\label{h180}
\end{align}
where from now on we drop the redundant symbol $\abs{\zeta_p}$ whenever $\zeta_p = \abs{\zeta_p}$. Substituting Eqs. (\ref{f150}-\ref{f180}) into Eqs. (\ref{f120}-\ref{f130}) one obtains
\begin{align}
H_L = 2 \int d \vp \; \zeta_p^2 \Bigl( \abs{c_{+,1}(\vp)}^2 +  \abs{c_{+,2}(\vp)}^2 \Bigr) \Theta(k_0^2 - p^2) ,
\label{h190}
\end{align}
and
\begin{align}
H_D = -2 \int d \vp \, \abs{\zeta_p}^2 \Bigl(c_{-,1}^*(\vp)c_{-,2}(\vp)  +  c_{-,2}^*(\vp)c_{-,1}(\vp)\Bigr) \Theta(p^2 - k_0^2) .
\label{h200}
\end{align}
Since $c_{-,1}^*c_{-,2}  +  c_{-,1}c_{-,2}^* = 2 \operatorname{Re} c_{-,1}^*c_{-,2}$ has not a definite sign and can take any value, it is clear that $H_D$ is not bounded from below.

\subsection{Recovering the angular spectrum representation}

To make a connection with the \emph{angular spectrum} theory in classical optics \cite{MandelBook}, let us begin by remarking that in free space, the diverging and converging exponential functions in Eq. \eqref{h160} cannot  \emph{both} represent physical solutions of Eq. \eqref{h140} for $z$ either positive or negative, since unbounded exponentially growing functions cannot belong to the spectrum of a realistic physical theory. Therefore, we must require that
\begin{align}
\begin{cases}
	c_{-,2}(\vp)=0, \qquad z >0, \\
  c_{-,1}(\vp)=0, \qquad z \leq 0. 
\end{cases}
\label{h210}
\end{align}
However, this means that $c_{-,1}^*(\vp)c_{-,2}(\vp) =0$ \emph{everywhere} and, consequently, $H_D = 0$. Therefore, the ``dark'' component of the Hamiltonian does not contribute to the \emph{physical} energy of the field. 

As shown with wealth of details by Mandel and Wolf in \cite{MandelBook}, the angular spectrum of a wavefield is \emph{uniquely} determined in the half-space $z \geq 0$ ($z \leq 0$) for well-behaving forward (backward) propagating fields. Let us fix, from now on,  $z \geq 0$ for the sake of definiteness.  In order to recover standard the angular spectrum from our Eqs. (\ref{h150}-\ref{h160}) we have to choose $c_{+,1}(\vp) = c_{-,1}(\vp) \equiv a(\vp)$ and $c_{\pm,2}(\vp) =0$, where the amplitude $a(\vp)$ is the quantity customarily dubbed ``angular spectrum''. In this case we can rewrite the original field $\psi(\vx,z)$ in the following form:
\begin{align}
\psi(\vx,z \geq 0) = & \; \frac{1}{2 \pi} \int d \vp \, Q(\vp,z) e^{i \vp \cdot \vx} \nonumber \\
 = & \; \frac{1}{2 \pi} \int d \vp \, e^{i \vp \cdot \vx}  \Bigl[ Q_+(\vp,z)\Theta(k_0^2 - p^2) +  Q_-(\vp,z)\Theta(p^2 - k_0^2  )\Bigl] \nonumber \\
 = & \; \frac{1}{2 \pi} \int d \vp \, a(\vp) \, e^{i \zeta_p z} \, e^{i \vp \cdot \vx} ,
\label{h220}
\end{align}
where the last equality follows from the fact that 
\begin{align}
\begin{cases}
e^{i \zeta_p z} = e^{i \abs{\zeta_p} z}, \qquad p^2 \leq k_0^2, \\
e^{i \zeta_p z} = e^{- \abs{\zeta_p} z}, \qquad p^2 > k_0^2, 
\end{cases}
\label{h230}
\end{align}
where $\zeta_p = + (k_0^2 - p^2)^{1/2}$. 
%
Substituting Eq. \eqref{h220} into Eq. \eqref{h40} and integrating over the $xy$-plane,  one obtains  
\begin{align}
H = & \; \int d \vx \, \mathscr{H} \nonumber \\
 = & \; 2 \int d \vp \, \zeta_p \operatorname{Re} \zeta_p \, e^{-2 z \operatorname{Im} \zeta_p }\abs{a(\vp)}^2 \nonumber \\
 = & \; 2 \int d \vp \, \abs{a(\vp)}^2 \zeta_p^2 \, \Theta\left(k_0^2 - p^2 \right),
\label{h2800}
\end{align}
where $\zeta_p^2 \, \Theta\left(k_0^2 - p^2 \right) \geq 0$ by definition.
This Hamiltonian is manifestly positive semidefinite and, therefore, physically acceptable. Moreover, Eq. \eqref{h2800} trivially implies that
\begin{align}
\pd{H}{z} =0,
\label{h2900}
\end{align}
which shows that energy is conserved during free propagation of wavefields represented in the angular spectrum form.

\subsubsection{Effective-Lagrangian theory for the angular spectrum}

Comparing the first and the last lines of Eq. \eqref{h220} one sees that 
\begin{align}
Q(\vp,z) = a(\vp)e^{i \zeta_p z}  \qquad \Rightarrow \qquad a(\vp) = Q(\vp,0) .
\label{h240}
\end{align}
However, by definition $Q(\vp,0)$ is the inverse Fourier transform of $\psi(\vx,0)$. Therefore, the angular spectrum $a(\vp)$ is \emph{uniquely} determined by the knowledge of $\psi(\vx,0)$ solely. This seems to be in contradiction with  Eq. \eqref{h110} which is a \emph{second-order} differential equation whose uniqueness of the solution require the knowledge of both 
\begin{align}
\left. Q(\vp,z) \right|_{z=0} \qquad \text{and} \qquad \left.\pd{Q(\vp,z)}{z}\right|_{z=0} =  P^*(\vp,0).
\label{h250}
\end{align}
%
The clarification of the foregoing apparent paradox lies in the observation that if we derive with respect to $z$ both sides of the leftmost part of Eq. \eqref{h240}, we obtain
\begin{align}
\pd{Q(\vp,z)}{z} = i \zeta_p Q(\vp,z),
\label{h260}
\end{align}
which is a \emph{first-order} differential equation. For reasons that will be soon clear, let us define 
\begin{align}
Q(\vp,z) = A_+^*(\vp,z)\Theta(k_0^2-p^2) + A_-(\vp,z) \Theta(p^2-k_0^2),
\label{h270}
\end{align}
where the presence of $A_-$ instead of $A_-^*$ in the equation above is not accidental.
Then, Eq. \eqref{h260} yields to two independent first-order equations of motion
\begin{align}
\pd{}{z}A_+(\vp,z) = -i \zeta_p A_+(\vp,z), \qquad \text{and} \qquad  \pd{}{z}A_-(\vp,z) = - \abs{\zeta_p} A_-(\vp,z),
\label{h280}
\end{align}
which can be derived from the \emph{effective first-order Lagrangians} $\mL_+$ and $\mL_-$  defined as
\begin{align}
\mL_+ = -\frac{i}{2} \left( A_+ \partial_z A_+^* -  A_+^* \partial_z A_+ \right) - \zeta_p A_+^* A_+, \qquad \text{for} \qquad \zeta_p = \zeta_p^*,
\label{h290}
\end{align}
and 
\begin{align}
\mL_- = \frac{1}{2} \left( A_- \partial_z  {\bar{A}_-} -  \bar{A}_- \partial_z A_- \right) - \abs{\zeta_p} \bar{A}_- A_-, \qquad \text{for} \qquad \zeta_p = -\zeta_p^*.
\label{h300}
\end{align}
From the latter equation follows that $\bar{A}_{-}(\vp,z)$ obeys the equation of motion 
\begin{align}
  \pd{}{z}{\bar{A}_-}(\vp,z) =  \abs{\zeta_p} {\bar{A}_-}(\vp,z).
\label{h310}
\end{align}
Therefore, $\bar{A}_{-}(\vp,z) \neq A_-^*(\vp,z)$ and it must be regarded as an \emph{independent variable}. If we choose both variables  $A_{-}(\vp,z)$ and $\bar{A}_{-}(\vp,z)$ \emph{reals}, then also $\mL_- $ is real and it becomes a physically admissible Lagrangian. However, while the Hamiltonian density
\begin{align}
\mathscr{H}_+ = \zeta_p \abs{ A_+}^2
\label{h320}
\end{align}
is positive semidefinite, the same is not true for 
\begin{align}
\mathscr{H}_- = \abs{\zeta_p} \bar{A}_{-} A_-,
\label{h330}
\end{align}
because $\bar{A}_{-}$ and $A_-$ can take, \emph{independently}, any real value.

Last but not least, it should be noticed that the solutions of Eq. \eqref{h310} are, evidently, exponential functions diverging for $z \to \infty$:
\begin{align}
{\bar{A}_-}(\vp,z) =  {\bar{A}_-}(\vp,0) \, e^{\abs{\zeta_p}z}.
\label{h340}
\end{align}
This is really curious: Although we had  removed \emph{ad hoc} such ``unstable'' solutions from the expression \eqref{h220} of the field, they entered back in the game to ensure the existence of a proper Lagrangian $\mL_-$. Indeed, if in Eq. \eqref{h300} one replaced ${\bar{A}_-}$ with ${A_-^*}$, the corresponding Lagrangian would become complex.

\section{Helmholtz equation in Dirac form}

In the foregoing section we have seen that the angular spectrum of a field obeys a first-order differential equation. However, this is not the only way to reduce the Helmholtz equation to a first-order form. To show this, let us rewrite the HWE in the compact form
\begin{align}
\pdd{\psi}{z}= - \left( \nabla^2 + k_0^2 \right) \psi,
\label{d10}
\end{align}
where, as usual in this work, $\nabla^2 = \partial_x^2 + \partial_y^2$.
A \emph{formal} solution of this equation can be written in an operator form as
\begin{align}
\psi(\vx,z) = e^{i z \sqrt{\nabla^2 + k_0^2}} \, \psi(\vx,0).
\label{d20}
\end{align}
Here, the ``Hamiltonian'' differential operator 
\begin{align}
H = \sqrt{\nabla^2 + k_0^2},
\label{d30}
\end{align}
is reminiscent of the Hamiltonian of a relativistic free particle $H = \sqrt{p^2 c^2 + m^2 c^4}$. The problems with our $H$ in Eq. \eqref{d30} are the same ones encountered in quantum mechanics when extending the Sch\"{o}dinger equation to the relativistic regime. The square root operator on the right in Eq. \eqref{d30} contains all powers of the $\nabla$ operator. Even worst, in our case $\nabla^2 \sim -p^2$. Therefore,  $\sqrt{-p^2 + k_0^2}$  even becomes purely imaginary for $p^2>k_0^2$ hence, apparently, breaking down the unitarity of the theory. In quantum mechanics this problem was brilliantly solved by Dirac who managed to reduce the second-order differential equation in Eq. \eqref{d10} to a first-order form, without altering the physics of the problem. 

Following in Dirac's footsteps, first we write a first-order equation of the form
\begin{align}
\pd{\psi}{z} = \alpha_1 \pd{\psi}{x} + \alpha_2 \pd{\psi}{y} + \beta k_0^2 \psi,
\label{d40}
\end{align}
and then we try to determine the unknown coefficients $\alpha_i, \beta$ by imposing that $\psi$ must also satisfy the second-order HWE. Therefore, iterating Eq. \eqref{d40}, we find
\begin{align}
\pdd{\psi}{z} = & \;  \alpha_1^2 \pdd{\psi}{x} + \alpha_2^2 \pdd{\psi}{y} +\left( \alpha_1 \alpha_2 + \alpha_2 \alpha_1\right) \frac{\partial^2 \psi}{\partial x \partial y} \nonumber \\
 & \;  + k_0^2 \left( \alpha_1 \beta + \beta \alpha_1\right) \frac{\partial \psi}{\partial x } + k_0^2 \left( \alpha_2 \beta + \beta \alpha_2\right) \frac{\partial \psi}{\partial y } + \beta^2 k_0^4 \psi \nonumber \\
\equiv  & \; -\pdd{\psi}{x} - \pdd{\psi}{y} -k_0^2 \psi.
\label{d50}
\end{align}
The first two and the last lines of Eq. \eqref{d50} coincide providing that
\begin{equation}
\begin{split}
 \alpha_1^2 = \alpha_2^2 = & \; -1, \nonumber \\
 \alpha_1 \alpha_2 + \alpha_2 \alpha_1 = & \; 0, \nonumber \\
 \alpha_i \beta + \beta \alpha_i = & \; 0, \nonumber \\
 \beta^2 k0^2= & \; -1.
\end{split}
\label{d60}
\end{equation}
It is not difficult to verify that the equations above are satisfied by choosing
\begin{align}
\alpha_1 = i \sigma_1 = 
\begin{pmatrix}
0 & i \\
i & 0
\end{pmatrix}, \qquad \alpha_2 = i \sigma_2 = 
\begin{pmatrix}
0 & 1 \\
-1 & 0
\end{pmatrix}, \qquad \beta k_0 = i \sigma_3 = 
\begin{pmatrix}
i & 0 \\
0 & -i
\end{pmatrix},
\label{d70}
\end{align}
where $\sigma_\mu$ are the Pauli matrices. Differently from the Dirac's theory, here all the $\alpha$s and $\beta$ matrices are \emph{anti-Hermitean}. Then, we can rewrite Eq. \eqref{d40} as
\begin{align}
\pd{}{z}\psi(\vx,z) = & \;  i \left( \sigma_1 \pd{}{x} + \sigma_2 \pd{}{y} + \sigma_3\, k_0 \right) \psi(\vx,z) 
%
,
\label{d80}
\end{align}
where the original scalar field $\psi(\vx,z)$ must be now regarded as a doublet 
\begin{align}
\psi(\vx,z) = \begin{pmatrix}
\Psi_1(\vx,z) \\
\Psi_2(\vx,z)
\end{pmatrix}.
\label{d90}
\end{align}
The appearance of the two functions $\Psi_i$ instead of the unique original one $\psi$, may be hardly surprising as clearly explained by Messiah \cite{MessiahBook}. Indeed, the solution of a second-order differential equation with respect to $z$, as the HWE is, requires the knowledge of both $\psi(\vx,z)$ and $\partial_z\psi(\vx,z)$ evaluated at the initial position $z=0$. Therefore, converting the second-order HWE to a first-order Dirac form without loosing information, necessarily introduces a two-component wavefield.

In order to find a Lagrangian for Eq. \eqref{d80} it is convenient first to rewrite it in the more suggestive form
\begin{align}
\bigl( i \gamma^\mu \partial_\mu + k_0 \bigr)\psi =0,
\label{d100}
\end{align}
where the three $2 \times 2$  matrices $\gamma^\mu$ are defined as:
\begin{align}
\gamma^1 = - i \sigma_2, \qquad \gamma^2 =  i \sigma_1, \qquad \gamma^3 =  \sigma_3.
\label{d110}
\end{align}
Then, if we define $\bar{\psi} \equiv \psi^\dagger \gamma^3$, with $(\psi^\dagger)_i = \Psi_i^*$, we can straightforwardly write down the Lagrangian as
\begin{align}
\mL = \bar{\psi} \bigl( i \gamma^\mu \partial_\mu - k_0 \bigr)\psi.
\label{d120}
\end{align}

\section{Conclusions}

This is the first part of a work in progress. It serves to establish the basic formalism before proceeding with the ``phenomenological'' quantization of both Helmholtz and paraxial wave equation, which will be presented in part II.

%
%

%
\end{document}